# Effect of capping layer on spin-orbit torques

Chi Sun[1], Zhuo Bin Siu[1], Seng Ghee Tan[2], Hyunsoo Yang[1] and Mansoor B. A. Jalil[1,a)]

[1] Department of Electrical and Computer Engineering, National University of Singapore, Singapore, 117576, Singapore
[2] Department of Physics, National Taiwan University, Taipei, 10617, Taiwan

**ABSTRACT**

In order to enhance the magnitude of spin-orbit torque (SOT), considerable experimental works have been devoted to studying the thickness dependence of the different layers in multilayers consisting of heavy metal (HM), ferromagnet (FM) and capping layers. Here we present a theoretical model based on the spin-drift-diffusion (SDD) formalism to investigate the effect of the capping layer properties such as its thickness on the SOT observed in experiments. It is found that the spin Hall-induced SOT can be significantly enhanced by incorporating a capping layer with opposite spin Hall angle to that of the HM layer. The spin Hall torque can be maximized by tuning the capping layer thickness. However, in the absence of the spin Hall effect (SHE) in the capping layer, the torque decreases monotonically with capping layer thickness. Conversely, the spin Hall torque is found to decrease monotonically with the FM layer thickness, irrespective of the presence or absence of SHE in the capping layer. All these trends are in correspondence with experimental observations. Finally, our model suggests that capping layers with long spin diffusion length and high resistivity would also enhance the spin Hall torque.

## I. INTRODUCTION

Spin-orbit torques (SOTs) have recently become one of the most active research topic in spintronics due to their prospective applications in realizing current-induced magnetization switching (CIMS)[1,2], persistent magnetization oscillation[3] and fast domain wall motion[4,5]. SOT is conventionally induced by applying an in-plane current

---

[a)] Electronic address: elembaj@nus.edu.sg

through heavy metal (HM)/ ferromagnet (FM)/ Cap multilayers. In a HM with strong spin-orbit coupling, electrons are spin polarized and subsequently transfer angular momentum to the adjacent FM layer. Thus, they exert SOT on the FM, which can switch the magnetization. There are at least two mechanisms which can contribute to this phenomenon, namely the bulk spin Hall effect, and the interfacial Rashba effect.[6,7]

In order to enhance the magnitude of SOT, considerable experimental works have been devoted to studying the effects of using different HM materials replacement, varying the HM and FM layer thicknesses, as well as the physical attributes of the capping layer.[8-11] Commonly used capping layer materials include Cu and oxides such as MgO. In these structures, the capping layer merely performs a protective function. It has also been reported that SOT can be enhanced experimentally by using another HM as a capping layer in which the two HMs possess opposite spin Hall angles.[11,12] Among HM materials, Pt and Ta have been widely used due to their relatively high positive and negative spin Hall angles, respectively.[1,8,13] Recently, W and Hf have also been studied as negative spin Hall materials.[9,11,14,15] It was observed that the SOT can be optimized by varying the different layer thicknesses of the trilayer structure and engineering the physical properties of the capping layer, an optimized SOT can be achieved.[10,11]

To explain these experimental results and understand the underlying physics, we develop a model based on the spin-drift-diffusion (SDD) theory. Specifically, the model would be used to study the effect of capping layer on SOTs in perpendicularly magnetized HM/FM/Cap systems (Fig. 1). Here we consider two cases: i) a capping layer which acts only as a protective layer, e.g., one made of normal metal (NM), and ii) a capping layer made of HM material, that provides a spin polarized current source via the spin Hall effect. For case i), we find that SOT decreases with increasing capping layer thickness ($t_{Cap}$), in agreement with the experimental results from Ref. 10 where Cu is used as the capping layer. While for case ii), the calculations reveal a maximum value of SOT, as has been realized in experiments.[11] We also investigate the HM and FM layer thickness dependence of the SOT, and obtain the same trends reported in previous experimental works.[8-11] In addition to the thickness dependence, we also assess how other physical properties of the capping layer such as the spin diffusion length and resistivity affect the SOT, and suggest a means to enhance it. However, the drift-diffusion approach being used in our present model does not capture the effect of interfacial spin-orbit coupling, i.e., we focus only on the bulk spin Hall effect in the absence of interfacial Rashba effect.

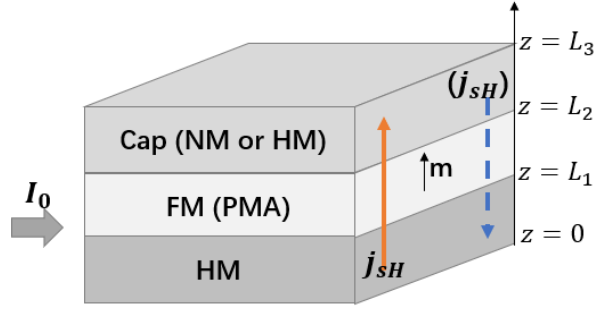

FIG. 1. Schematic diagram of the HM/FM/Cap multilayers in this work. Here the FM possesses perpendicular magnetic anisotropy (PMA), and we assume the local magnetization direction is perpendicular to plane (i.e. along the $z$ direction) in the FM. The charge current $I_0$ flows in-plane and introduces spin polarized currents due to the spin Hall effect (SHE) in the HM. Here we will focus on perpendicular-to-plane transport along the $z$ axis.

## II. THEORY AND MODEL

The spin-drift-diffusion (SDD) approach of Valet and Fert[16] is based on integrating the Boltzmann equation to derive transport equations that relate the charge and spin densities on the corresponding currents. The diffusion equation for arbitrary spin quantization axis is described by a $(2 \times 2)$ matrix equation in spin space,[17,18]

$$\widehat{D} \cdot \frac{\partial^2 \hat{\mu}}{\partial z^2} = \frac{1}{\tau_{sf}} \left[ \hat{\mu} - \hat{1} \frac{Tr\{\hat{\mu}\}}{2} \right], \quad (1)$$

where $\widehat{D}$ is the spin diffusion matrix, $\hat{1}$ is the unit matrix, $\hat{\mu}$ is the electrochemical potential and $\tau_{sf}$ is the spin-flip relaxation time. The corresponding charge and spin current equations are given by

$$\hat{j} = -\widehat{D} \cdot \frac{\partial \hat{\mu}}{\partial z}, \quad (2)$$

where $\hat{j}$ is the current matrix, which incorporates the usual charge current $j_c$ and spin current $j_s$ via the relation: $\hat{j} = \frac{1}{2}(j_c\hat{1} + j_s\hat{\sigma})$. We now consider the solutions to $\hat{\mu}$ and $\hat{j}$ in Eqs. (1) and (2) within the FM and NM layers shown schematically in Fig. 1, respectively.

## A. Spin currents and spin accumulation in FM layer

In the FM layer, electron spins are aligned along the local magnetization direction due to a strong exchange field over a length scale much smaller than the mean free path or the spin diffusion length. The spin current polarized transverse to the magnetization direction quickly dissipates due to a precession-induced dephasing of spins. In other words, the transverse components (i.e., $x$ and $y$ components) of spin current are much smaller than that polarized along the local magnetization direction (i.e., $z$ direction). Since the spin reference axis is defined along the local magnetization direction, the off-diagonal terms representing the transverse spin components in $\hat{\mu}$ and $\hat{j}$ can be treated as negligible within the FM layer. Considering its longitudinal component along the local magnetization direction (i.e., diagonal terms in the matrix) first, we have the usual collinear equation:

$$D_{\uparrow(\downarrow)} \frac{\partial^2 \mu_{\uparrow(\downarrow)}}{\partial z^2} = \frac{1}{2\tau_{sf}} [\mu_{\uparrow(\downarrow)} - \mu_{\downarrow(\uparrow)}], \tag{3}$$

where $\uparrow$ and $\downarrow$ denote spin up and down. Eq. (3) can be rearranged to yield

$$\frac{\partial^2 (\mu_\uparrow - \mu_\downarrow)}{\partial z^2} = \frac{1}{(l_{sf}^L)^2} (\mu_\uparrow - \mu_\downarrow), \tag{4}$$

$$\frac{\partial^2 (\mu_\uparrow + \mu_\downarrow)}{\partial z^2} = \eta \frac{\partial^2 (\mu_\uparrow - \mu_\downarrow)}{\partial z^2}, \tag{5}$$

where $l_{sf}^L$ is the longitudinal spin diffusion length which is defined as $\frac{1}{\left(l_{sf}^L\right)^2} = (\frac{1}{l_\uparrow^2} + \frac{1}{l_\downarrow^2})/2$ with $l_\uparrow^2 = D_\uparrow \tau_{sf}$ and $l_\downarrow^2 = D_\downarrow \tau_{sf}$, and $\eta$ is the spin asymmetry of the diffusion coefficients, i.e. $\eta = (D_\downarrow - D_\uparrow)/(D_\downarrow + D_\uparrow)$. Eq. (4) is equivalent to $\frac{\partial^2 \Delta\mu}{\partial z^2} = \frac{\Delta\mu}{l_{sf}^2}$ derived from the Boltzmann equation by Valet and Fert.[16] The solutions of Eqs. (4) and (5) are

$$\mu_\uparrow = (\eta + 1)\left[A \exp\left(\frac{z}{l_{sf}^L}\right) + B \exp\left(\frac{-z}{l_{sf}^L}\right)\right] + Cz + G, \tag{6}$$

$$\mu_\downarrow = (\eta - 1)\left[A\ \exp\left(\frac{z}{l_{sf}^L}\right) + B\ \exp\left(\frac{-z}{l_{sf}^L}\right)\right] + Cz + G, \quad (7)$$

where $A$, $B$, $C$ and $G$ are coefficients to be determined by boundary conditions which will be discussed later. Charge and longitudinal spin chemical potential can be defined as $\mu_c = (\mu_\uparrow + \mu_\downarrow)/2$ and $\mu_s^L = (\mu_\uparrow - \mu_\downarrow)/2$, respectively. From Eqs. (6) and (7), the expressions of these are given by

$$\mu_c = \eta\left[A\ \exp\left(\frac{z}{l_{sf}^L}\right) + B\ \exp\left(\frac{-z}{l_{sf}^L}\right)\right] + Cz + G, \quad (8)$$

$$\mu_s^L = A\ \exp\left(\frac{z}{l_{sf}^L}\right) + B\ \exp\left(\frac{-z}{l_{sf}^L}\right). \quad (9)$$

Similarly, the transverse components of spin chemical potential $\mu_s^T$ take the same form as Eq. (9), but with a different value of transverse diffusion length $l_{sf}^T$.[19,20] Here we introduce $l_{sf}^T$ in a diffusive manner to describe the indeed dephasing process. This dephasing happens over a distance much smaller than the longitudinal spin diffusion length $l_{sf}^L$ in normal transition metal ferromagnets and their alloys.[18,21,22] Thus, $l_{sf}^T$ is introduced as an input parameter which is one order of magnitude smaller than $l_{sf}^L$ in our model. This approximation also ensures spin transverse components can be treated as negligible compared to that along the local magnetization direction.[6,20] The corresponding charge and spin currents can be readily obtained by substituting Eqs. (8) and (9) into Eq. (2). We then obtain

$$j_c = -C(D_\uparrow + D_\downarrow), \quad (10)$$

$$j_s^L = -C(D_\uparrow - D_\downarrow) - \frac{2\tilde{D}}{l_{sf}^L}[A\ \exp\left(\frac{z}{l_{sf}^L}\right) - B\ \exp\left(\frac{-z}{l_{sf}^L}\right)], \quad (11)$$

where $\tilde{D} = 2D_\uparrow D_\downarrow/(D_\uparrow + D_\downarrow)$. As for transverse spin currents $j_s^T$, only the second term in Eq. (11) exists with a finite $l_{sf}^T$,[20,23,24] i.e., $j_s^T = -\frac{2\tilde{D}}{l_{sf}^T}[A\ \exp\left(\frac{z}{l_{sf}^T}\right) - B\ \exp\left(\frac{-z}{l_{sf}^T}\right)]$. Actually, since we only consider the transport in the vertical $z$ direction under an applied charge current in the in-plane ($x - y$ plane) direction, there is zero net charge current flow in the $z$ direction, as expressed in Eq. (10). This assumption holds for all values of $z$, i.e., for all layers in the whole structure.

## B. Spin currents and spin accumulation in HM/Cap layer

In the normal metal (NM), the spin and charge chemical potential satisfy the following drift-diffusion equations

$$\frac{\partial^2 \mu_c}{\partial z^2} = 0, \tag{12}$$

$$\frac{\partial^2 \mu_s}{\partial z^2} = \frac{\mu_s}{l_{sf}^2}. \tag{13}$$

The solutions of Eqs. (12) and (13) are

$$\mu_c = -Cz + G, \tag{14}$$

$$\mu_s = A \exp\left(\frac{z}{l_{sf}}\right) + B \exp\left(\frac{-z}{l_{sf}}\right). \tag{15}$$

The above equations correspond to Eqs. (8) and (9) for the FM layer, in the limit of $\eta = 0$ due to $D_\uparrow = D_\downarrow$ in the NM. Note that in the NM layer, the expressions for longitudinal and transverse spin components take the same form, and therefore we do not differentiate between them. In our system, we treat the heavy metal (HM) layer as a NM layer exhibiting a spin Hall effect, so that the spin current in HM consists of diffusion and spin Hall drift contributions. Thus, Eq. (2) is modified to $j_s = -D\frac{\partial \mu_s}{\partial z} + j_{sH}$, where $j_{sH}$ is the bare spin Hall current generated directly by the SHE from the applied in-plane charge current $j_0$. In practical experiments, $j_0$ is applied to the whole stack and will be proportionately shunted into all three layers of the stack as described by

$$j_0^{HM} = \frac{1}{\rho_{HM}} \frac{t_{FM} + t_{HM} + t_{Cap}}{t_{FM}/\rho_{FM} + t_{HM}/\rho_{HM} + t_{Cap}/\rho_{Cap}} j_0, \tag{16}$$

$$j_0^{Cap} = \frac{1}{\rho_{Cap}} \frac{t_{FM} + t_{HM} + t_{Cap}}{t_{FM}/\rho_{FM} + t_{HM}/\rho_{HM} + t_{Cap}/\rho_{Cap}} j_0, \tag{17}$$

where $j_0^{HM(Cap)}$ is the in-plane charge current being shunted into HM (Cap) layer, and $t$ represents the thickness and $\rho$ the resistivity of each layer. The spin Hall current generated in a HM (Cap) layer by $j_0$ can be expressed as $j_{sH}^{HM(Cap)} = \theta_{sH}^{HM(Cap)} j_0^{HM(Cap)}$, where $\theta_{sH}^{HM(Cap)}$ is the so-called spin Hall angle of HM (Cap) layer. Here, we treat the values of the spin Hall angles as input parameters, which can be extracted from experiments. For comparison, we also consider the case of a capping layer without spin Hall effect, i.e., $\theta_{SH}^{Cap} = 0$. Having considered the magnitude of the spin Hall current, we now turn to its polarization direction. We assume a planar Hall effect (PHE), where $j_{sH}$ is polarized in an arbitrary in-plane $(x - y)$ direction with angle $\theta$ between $j_{sH}$ and the $x$ axis. Thus, as before, in the polarization direction of $j_{sH}$ we have: $j_{s,\|sH} = -D\frac{\partial \mu_{s,\|sH}}{\partial z} + j_{sH}$, while in the perpendicular direction current: $j_{s,\perp sH} = -D\frac{\partial \mu_{s,\perp sH}}{\partial z}$. The spin current expressions in the $x$ and $y$ directions are then given by

$$j_{s,x} = j_{s,\|sH}\cos\theta - j_{s,\perp sH}\sin\theta, \tag{18}$$

$$j_{s,y} = j_{s,\|sH}\sin\theta + j_{s,\perp sH}\cos\theta. \tag{19}$$

Similar transformations are also applied to spin chemical potential $\mu_s$.

### C. Boundary conditions and spin Hall torque

In order to solve the unknown set of coefficients in the expressions given above, appropriate boundary conditions need to be implemented. The first set of boundary conditions comes from zero spin accumulation $\mu_s = 0$ at the terminals of the device. It is assumed that semi-infinite non-magnetic metal leads are attached to both ends of the system at $z = L_0$ and $L_3$ (see Fig. 1). It has been pointed this approximation is consistent with a metal contact with infinite conductivity,[18-20] which can be achieved by having a NM lead with a much larger cross-sectional area than the submicrometre-sized region of the device. Thus, it is reasonable to apply the zero spin accumulation boundary condition at the terminals. Additionally, we also set the electrostatic potential to be grounded at one end, i.e., $\mu_c(z = L_3) = 0$.

The second set of boundary conditions are related to electrochemical potentials and charge and spin currents, which are determined by the spin-dependent interfacial conductance at $L_1$ and $L_2$, i.e., $G_\uparrow$, $G_\downarrow$ and $G_{\uparrow\downarrow}$. $G_{\uparrow(\downarrow)}$ is the interfacial conductance experienced by majority (minority) spin electrons polarized along the local magnetization $z$ direction, while $G_{\uparrow\downarrow}$ represents the spin mixing conductance that couples the two transverse (i.e. $x$ and $y$) components of the spin accumulation across the interface. $G_{\uparrow(\downarrow)}$ may be characterized by a spin asymmetry factor $\gamma$, i.e., $G_\uparrow = G_0(1+\gamma)/2$ and $G_\downarrow = G_0(1-\gamma)/2$. $G_{\uparrow(\downarrow)}$ relates the charge and spin currents to the discontinuity in the electrochemical potentials across the interfaces $z = L_1$ and $L_2$, such that the boundary conditions are given by[18,25]

$$j_c|_{z=L_1,L_2} = (G_\uparrow + G_\downarrow)\Delta\mu_c|_{z=L_1,L_2} + (G_\uparrow - G_\downarrow)\Delta\mu_s|_{z=L_1,L_2}, \quad (20)$$

$$j_{s,z}|_{z=L_1,L_2} = (G_\uparrow - G_\downarrow)\Delta\mu_c|_{z=L_1,L_2} + (G_\uparrow + G_\downarrow)\Delta\mu_s|_{z=L_1,L_2}, \quad (21)$$

where $\Delta\mu_c$ is the charge chemical potential discontinuity across the interface, and similarly for $\Delta\mu_s$. The spin mixing conductance $G_{\uparrow\downarrow}$, includes both reflected mixing conductance $G_{\uparrow\downarrow}^r$ and transmitted mixing conductance $G_{\uparrow\downarrow}^t$. The transverse spin currents across the interfaces $z = L_1$ and $L_2$ can be expressed through the reflection and transmission matrices.[25] The derivation of these boundary conditions is given in the Appendix. From these boundary conditions, the transverse spin currents at the interfaces are given by[10]

$$j_{s,y}|_{z=L_1} = -2\text{Im}[G_{\uparrow\downarrow}^t]\mu_{s,x}|_{z=L_1^+} - 2\text{Re}[G_{\uparrow\downarrow}^t]\mu_{s,y}|_{z=L_1^+} + 2\text{Im}[G_{\uparrow\downarrow}^r]\mu_{s,x}|_{z=L_1^-} + 2\text{Re}[G_{\uparrow\downarrow}^r]\mu_{s,y}|_{z=L_1^-}, \quad (22)$$

$$j_{s,x}|_{z=L_1} = 2\text{Im}[G_{\uparrow\downarrow}^t]\mu_{s,y}|_{z=L_1^+} - 2\text{Re}[G_{\uparrow\downarrow}^t]\mu_{s,x}|_{z=L_1^+} - 2\text{Im}[G_{\uparrow\downarrow}^r]\mu_{s,y}|_{z=L_1^-} + 2\text{Re}[G_{\uparrow\downarrow}^r]\mu_{s,x}|_{z=L_1^-}, \quad (23)$$

$$j_{s,y}|_{z=L_2} = 2\text{Im}[G_{\uparrow\downarrow}^t]\mu_{s,x}|_{z=L_2^-} + 2\text{Re}[G_{\uparrow\downarrow}^t]\mu_{s,y}|_{z=L_2^-} - 2\text{Im}[G_{\uparrow\downarrow}^r]\mu_{s,x}|_{z=L_2^+} - 2\text{Re}[G_{\uparrow\downarrow}^r]\mu_{s,y}|_{z=L_2^+}, \quad (24)$$

$$j_{s,x}|_{z=L_2} = -2\text{Im}[G_{\uparrow\downarrow}^t]\mu_{s,y}|_{z=L_2^-} + 2\text{Re}[G_{\uparrow\downarrow}^t]\mu_{s,x}|_{z=L_2^-} + 2\text{Im}[G_{\uparrow\downarrow}^r]\mu_{s,y}|_{z=L_2^+} - 2\text{Re}[G_{\uparrow\downarrow}^r]\mu_{s,x}|_{z=L_2^+}. \quad (25)$$

Note that at the left hand side of the above four equations, we have already applied the continuity relations for transverse spin currents, e.g., $j_{s,y}|_{z=L_1^+} = j_{s,y}|_{z=L_1^-} = j_{s,y}|_{z=L_1}$. To simplify the calculation, the imaginary part of the mixing conductance can be neglected as it is approximately an order of magnitude smaller than the real part.[10] By implementing the above boundary conditions Eqs. (20) to (25), we can solve for the chemical potentials and currents throughout the HM/FM/Cap trilayer.

In this work, we aim to analyze the spin torque exerted to the FM layer. The spin Hall torque is defined as the difference between the transverse spin current at the two interfaces, which represents the absorbed transverse spin current in the FM layer,[10] i.e.,

$$\tau = j_{s,y}|_{z=L_1} - j_{s,y}|_{z=L_2}. \quad (26)$$

Note that in the absence of interfacial spin-flip scattering, spin transport transverse to the magnetization becomes decoupled from that longitudinal to it. As a result, only transverse (i.e. $x$ and $y$) components with their corresponding boundary conditions need to be solved in order to calculate the spin Hall torque defined in Eq. (26).

## III. RESULTS AND DISCUSSION

In our model, we consider two types of capping layer as mentioned before: i) capping layer that acts only as a protective layer, e.g. normal metal (NM); or ii) capping layer that acts as another heavy metal (HM) layer which serves as a spin-polarized current source via the spin Hall effect. Thus, we consider a trilayer comprising of a HM layer made of Pt, ferromagnetic layer (FM) of Co, and capping layer of Cu (NM) or W (HM) representing the above two cases, respectively. For our numerical calculations, we assume known experimental values for the bulk material parameters:[6,11,18,20] i) in HM layer, the parameter values are $\rho_{Pt} = 20\ \mu\Omega \cdot cm$ and $l_{sf}^{Pt} = 2.57$ nm; ii) in FM layer, $\rho_{Co} = 16\ \mu\Omega \cdot cm$, $l_{sf}^{L,Co} = 60$ nm, $l_{sf}^{T,Co} = 1$ nm and $\eta = 0.4$; iii) in NM capping layer, $\rho_{Cu} = 2.86\ \mu\Omega \cdot cm$ and $l_{sf}^{Cu} = 140$ nm; and iv) in HM capping layer, $\rho_W = 35\ \mu\Omega \cdot cm$ and $l_{sf}^W = 2.1$ nm. For the interfacial conductance, we assume the values used by Refs. 10 and 18 since there are no available experimental measurements, i.e., $G_\uparrow = 0.42 \times 10^{15}\ \Omega^{-1}m^{-2}$, $G_\downarrow = 0.36 \times 10^{15}\ \Omega^{-1}m^{-2}$, $Re[G_{\uparrow\downarrow}^r] = 0.5 \times 10^{15}\ \Omega^{-1}m^{-2}$ and $Re[G_{\uparrow\downarrow}^t] = 0.05 \times 10^{15}\ \Omega^{-1}m^{-2}$. As mentioned before, spin Hall current generated in the HM (Cap) layer can be described by the spin Hall angle. The spin Hall angles assumed for different materials are $\theta_{sH}^{Pt} = 0.1$, $\theta_{sH}^{Cu} = 0$ and $\theta_{sH}^{W} = -0.2$.[11,26]

### A. Layer thickness dependence of SOT

First, the effects of capping layer thickness on the spin Hall torque are investigated. We plot the spin Hall torque as a function of the capping layer thickness $t_{Cap}$ which is varied from 0 (no capping layer) to 10 nm for case i) $\theta_{sH}^{Cu} = 0$ and ii) $\theta_{sH}^{W} = -0.2$, respectively. Note that the magnitude of spin Hall torque has been normalized by the in-plane charge current density with $t_{Cap} = 0$ nm (i.e., a fixed charge current density). In Fig. 2(a), it is assumed in case i) that there is no spin Hall effect (i.e. $\theta_{Cap} = 0$) in the Cu capping layer. The spin Hall torque decreases with increasing $t_{Cap}$, which corresponds to the trend observed experimentally. In the inset of Fig. 2(a), we consider data from an experimental work (Ref. 10) and plot the SOT switching efficiency as a function of Cu capping layer thickness. The decreasing trend in the torque magnitude with increasing $t_{Cap}$ closely matches the theoretical results. However, as can be seen from Fig. 2(b), in the case of ii) capping layer of W which acts as another HM with an opposite spin Hall angle (i.e., $\theta_{Cap} < 0$), spin Hall torque achieves a maximum value at $t_{Cap} \sim 2$ nm, at which point the general decreasing trend starts to overcome the second SHE source from the capping layer. This behavior corresponds closely to that observed from another experimental work (Ref. 11), in which a torque maximum occurs at $t_{Cap} \sim 2$ nm, as shown in the inset of Fig. 2(b). Note that, as shown in the inset, the magnitude of SOT is calibrated by measuring the SOT effective field. Comparing cases i) and ii), besides the presence of a torque maximum, it can also be seen that the magnitude of spin Hall torque is generally enhanced by the presence of spin Hall effect in the capping layer. This is understandable since we have two HMs with opposite spin Hall angles, i.e., two sources which contribute to spin polarized currents of the same sign, thus enhancing the resultant spin Hall torque. However, as can be seen from the yellow dotted line in Fig. 2(a), if the capping layer is a HM with $\theta_{Cap} > 0$ (i.e. having the same sign as the other HM layer), it will compensate the spin Hall current generated in the other HM, resulting in a reduction of spin Hall torque. In this case, the capping layer thickness dependence shows a similar trend to the case when $\theta_{Cap} = 0$ (the blue solid line) in the same figure. The above results suggest that one could use a capping layer as another HM, but with an opposite spin Hall angle, in order to enhance the spin Hall torque and achieve a torque maximum at a certain capping layer thickness.

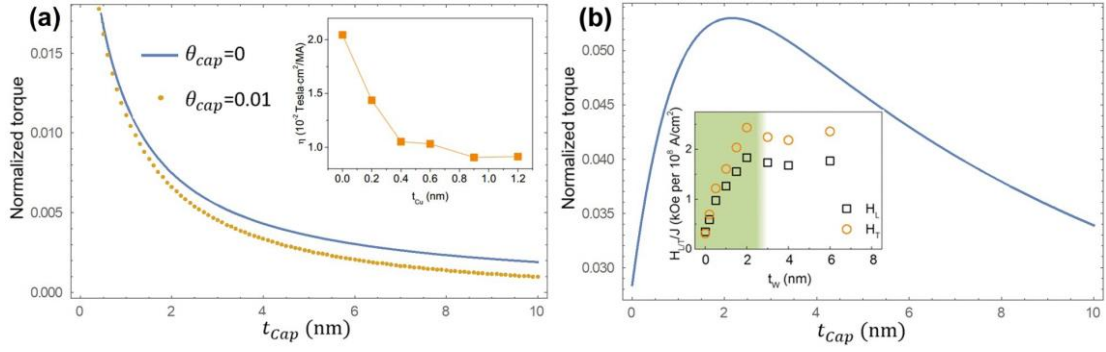

FIG. 2. The normalized torque as a function of capping layer thickness $t_{Cap}$ from 0 to 10 nm with (a) Cu as capping layer (i.e., $\theta_{Cap} = 0$), in which we also plot the case for $\theta_{Cap} > 0$ for comparison as shown by the yellow dotted line; (b) W as capping layer (i.e., $\theta_{Cap} < 0$). Here we set $t_{HM} = 4$ nm and $t_{FM} = 0.8$ nm. The insets are experimental results with corresponding capping material adapted from Refs. 10 and 11, respectively, which indicate similarity between the simulation and the experimental results.

Next, we study how the variation of FM layer thickness affects the magnitude of spin Hall torque (see Fig. 3). Similarly, the magnitude of spin Hall torque is normalized by the in-plane charge current density with $t_{Cap} = 2$nm and $t_{FM} = 0$ nm. Fig. 3(a) shows spin Hall torque as a function of $t_{FM}$ which is varied from 0 to 10 nm in case i) when using Cu as a capping material (i.e. $\theta_{Cap} = 0$). The spin Hall torque decreases with increasing $t_{FM}$ after achieving a peak at $t_{FM}$ is smaller than 1 nm. This trend corresponds to the experimental data from Ref. 10 as shown in the inset, where MgO and Ru were used as capping layer. MgO is a widely used capping layer with $\theta_{Cap} = 0$ while Ru possesses a negligible spin Hall angle.[27] In this experimental work, the measured quantity is the SOT switching efficiency which is a measure of the spin torque acting on the FM layer. Note that the experimental measurements do not start from $t_{FM} = 0$ nm but from $t_{FM} = 0.9$ nm because a finite thickness of the FM layer is needed to realize magnetization switching. As a result, the decreasing trend seen experimentally to some extent corresponds to our simulation results for $t_{FM}$ larger than 1 nm. Fig. 3(b) shows the result in case ii) when using W as a capping layer (i.e., $\theta_{Cap} < 0$), in which the trend of spin Hall torque is similar to case i). It can be also seen that the magnitude of spin Hall torque has been significantly enhanced in case ii). In general, the spin Hall torque shows a similar trend with $t_{FM}$, i.e., it decreases after achieving a maximum at a small thickness $t_{FM}$, regardless of whether the capping layer is a NM or HM. This peak value occurs at $t_{FM}$ that is comparable to $l_{sf}^T$ (i.e., transverse spin diffusion length of a FM), above which the absorption of transverse spin currents which gives rise to the spin torque starts to saturate.[10,28] In addition, by comparing with the yellow dotted line in Fig. 3(b), it can be seen a shorter $l_{sf}^T$ results in increased magnitude of spin Hall torque, which can be explained by the fact that a strong $s$-$d$ coupling leads to fast relaxation of transverse spin accumulation. This dependence of $l_{sf}^T$ also applies in the case of $\theta_{Cap} = 0$ shown in Fig. 3(a), and a more accurate value of $l_{sf}^T$ need to be experimentally determined to see if a better fit can be achieved with the faster decay trend seen in the inset experimental data (Ref. 10), based on a FM layer made of Co-Ni alloy.

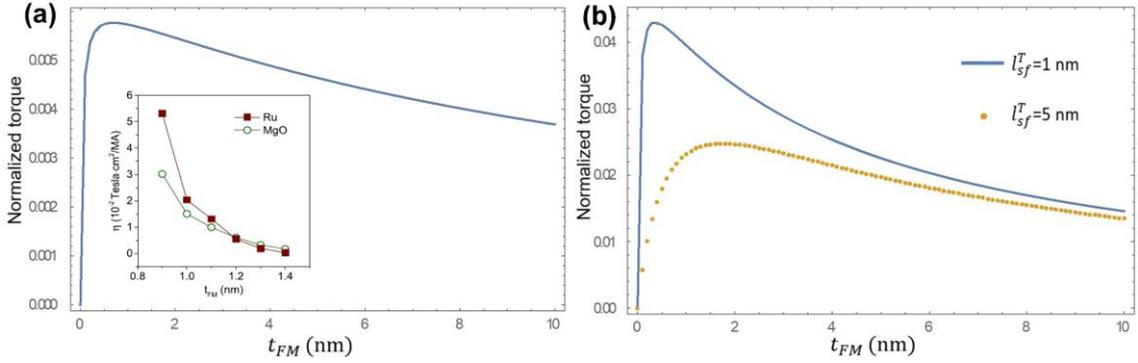

FIG. 3. The normalized torque as a function of FM layer thickness $t_{FM}$ from 0 to 10 nm with (a) Cu as capping layer (i.e., $\theta_{Cap} = 0$) and (b) W as capping layer (i.e., $\theta_{Cap} < 0$), for which we also plot the case of longer $l_{sf}^T$ (i.e., $l_{sf}^T = 5$ nm), shown by the yellow dotted line, for comparison. Here we set $t_{HM} = 4$ nm and $t_{Cap} = 2$ nm. The inset in (a) is the experimental result with $\theta_{Cap} = 0$ adapted from Ref. 10, which indicates the correspondence between the simulation and experimental trends for $t_{FM} \geq 1$ nm.

In addition, we investigate the HM thickness dependence of spin Hall torque. Here we only plot spin Hall torque as a function of $t_{HM}$ in case i), i.e., $\theta_{Cap} = 0$. This is because $t_{HM}$ dependence when $\theta_{Cap} < 0$ has already been discussed in Fig. 2. Besides, our previous calculations are based on $\theta_{HM} > 0$ since we had assumed Pt to be the HM layer. Now in order to compare with available experimental results using Hf which possesses a negative spin Hall angle as a HM layer, we modelled the case when $\theta_{HM} < 0$ in Fig. 4. The spin Hall torque increases in magnitude with increasing $t_{HM}$ from 0 to 10 nm with the torque saturating in magnitude at around $t_{HM} = 6$ nm. The results are in correspondence with comparable experimental results from Ref. 9 as shown in the inset of Fig. 4, which exhibits a similar trend. In the experiment of Ref. 9, the measured quantity is the SOT effective field which represents the spin Hall torque of our simulation. These results may be understood by noting that the relative amount of current flowing into the HM layer increases as its thickness is increased, thus generating a larger spin Hall current. However, the spin Hall current injection and hence the spin torque magnitude saturates as $t_{HM}$ approaches and exceeds the spin diffusion length of HM.[8,29,30] The similarity between our simulation and experiments also indicates that as $t_{HM}$ increases, the bulk spin Hall effect contributes to the observed saturation behavior of SOT. In addition, it has been indicated that the interfacial Rashba effect could significantly affect SOT at smaller $t_{HM}$.[9] However, the relative contributions of the bulk and interfacial spin torque still remain to be resolved.

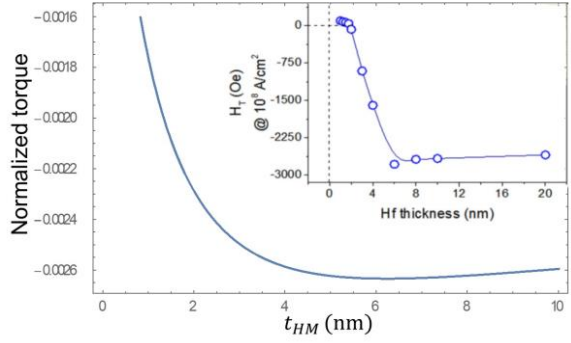

FIG. 4. The normalized torque as a function of HM layer thickness $t_{HM}$ from 0 to 10 nm with Cu as capping layer (i.e., $\theta_{Cap} = 0$). Here we set $t_{FM} = 0.8$ nm and $t_{Cap} = 2$ nm. The inset is the experimental results with $\theta_{HM} < 0$ and $\theta_{Cap} = 0$ from Ref. 9, thus indicating similarity between the simulation and experimental results.

## B. Capping layer dependence of SOT

Apart from layer thickness, we also consider how other physical properties of capping layer (i.e., spin diffusion length $l_{sf}^{Cap}$ and resistivity $\rho_{Cap}$) may affect the magnitude of spin Hall torque. In practice, $l_{sf}^{Cap}$ can be modulated by introducing impurities with strong spin-orbit coupling within the capping layer. We find that a strong spin depolarization in the capping layer (i.e. short $l_{sf}^{Cap}$) results in a reduction in spin Hall torque exerting to the FM layer for both case i) $\theta_{Cap} = 0$ and case ii) $\theta_{Cap} < 0$ shown in Fig. 5. This suggests that spin relaxation in the capping layer could have an effect in suppressing SOT in the FM layer. Comparing cases i) and ii), both give a similar trend of increasing SOT with $l_{sf}^{Cap}$, but with markedly different degrees of change. As the yellow dotted line shows, in the presence of SHE in the capping layer, i.e. case ii), the magnitude of spin Hall torque increases around 10% as $l_{sf}^{Cap}$ changes from 0.1 to 10 nm. However, for case i) plotted as the blue solid line, the relative increase in the magnitude of spin Hall torque is negligible (~ 0.02%). This suggests that one could use a capping layer material with long $l_{sf}^{Cap}$ to effectively enhance the spin Hall torque in the situation where there is a strong spin Hall effect in the capping layer [case ii)]. This effect significantly diminishes when there is no SHE in the capping layer.

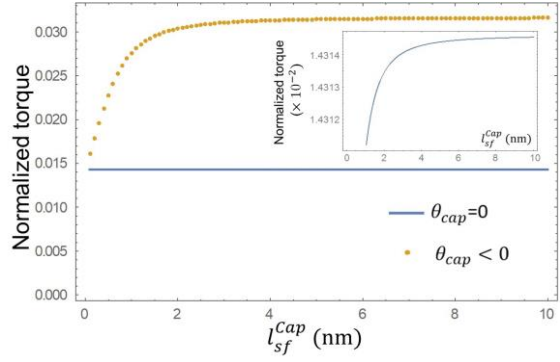

FIG. 5. The normalized torque as a function of capping spin diffusion length $l_{sf}^{Cap}$ from 0.1 nm to 10 nm with NM as capping layer (i.e., $\theta_{Cap} = 0$, blue solid line) and HM as capping layer (i.e., $\theta_{Cap} < 0$, yellow dotted line). The inset shows the magnified view of the data for $\theta_{Cap} = 0$.

Finally, we investigate the effect of capping resistivity $\rho_{Cap}$ on the spin Hall torque for case i) $\theta_{Cap} = 0$ and case ii) $\theta_{Cap} < 0$. In general, a high $\rho_{Cap}$ gives rise to a high spin Hall torque for both cases as shown in Fig. 6. Numerically, we found that a higher $\rho_{Cap}$ tends to increase the transverse spin accumulation at both interfaces $z = L_1$ and $L_2$, with both contributions resulting in increased torque exerted on the FM layer. This trend is also consistent with the previous studies on the capping layer effect on spin-valve structures in Refs. 18 and 19.

From the above results, we can make two observations:

a) The spin Hall torque can be enhanced significantly by the presence of SHE in the capping layer with an opposite spin Hall angle, in the case of which the spin Hall torque can be further optimized by tuning $t_{Cap}$ to coincide with the torque maxima, as shown in Fig. 2(b). Whether the capping layer possesses SHE or not gives rise to different layer thickness dependence of spin Hall torque. When there is no SHE in the capping layer ($\theta_{Cap} = 0$), spin Hall torque decreases monotonically with increasing $t_{Cap}$ in Fig. 2(a), without the torque maximum observed in the $\theta_{Cap} < 0$ case.

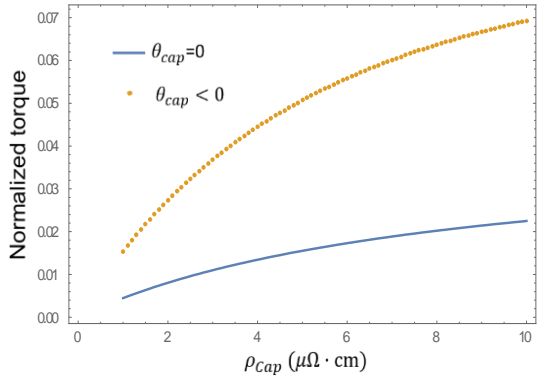

FIG. 6. The normalized torque as a function of capping spin resistivity $\rho_{Cap}$ from $1\,\mu\Omega\cdot\text{cm}$ to $10\,\mu\Omega\cdot\text{cm}$ with NM as capping layer (i.e., $\theta_{Cap} = 0$, blue solid line) and HM as capping layer (i.e., $\theta_{Cap} < 0$, yellow dotted line).

Meanwhile, spin Hall torque gives a similar decreasing trend with increasing $t_{FM}$ for both $\theta_{Cap} = 0$ and $\theta_{Cap} < 0$ (Fig. 3). The thickness dependence of the spin torque as predicted by our model bears close correspondence to available experimental results;[8-11]

b) The physical properties of the capping layer can affect the spin torque exerted on the FM layer. A capping layer with a long $l_{sf}^{Cap}$ and high resistivity tends to enhance the magnitude of spin Hall torque, leading to an increase in the overall SOT.

In closing, we comment on the limits of the applicability of our model which does not include the interfacial spin-orbit effects. It has been indicated that interfacial spin-orbit effects play an essential role in the analysis of spin transport in multilayers both experimentally[31,32] and theoretically[22,33-35], and contribute to spin accumulation and spin currents at the HM/FM interface. Based on Amin's SDD model[22,33] which has incorporated interfacial spin-orbit effects, our present model may be extended to include the interfacial spin-orbit effects by modifying the interfacial conductance values used in the boundary conditions by additional spin current terms, e.g., to account for spin memory loss due to spin-orbit coupling at the interface. However, further work needs to be done to determine the appropriate interfacial parameters in the Amin's model which are consistent with the parameters for bulk spin Hall effect in our model.

## IV. CONCLUSION

This study presents a theoretical model based on the spin-drift-diffusion (SDD) theory to study the capping layer effect on spin Hall torque. It is found that spin Hall torque can be significantly enhanced by using a capping layer with an opposite spin Hall angle to the HM. With this SHE in the capping layer, the capping layer thickness $t_{Cap}$ can be optimized to maximize the spin Hall torque. However, this torque maximum is absent for a capping layer with zero spin Hall angle (i.e., the capping layer acts only as a protective layer), for which the spin Hall torque decreases monotonically with increasing $t_{Cap}$. The predicted dependence of the spin torque on the layer thicknesses is in agreement with available experimental results. In addition to the layer thickness, we also analyzed the influence of $l_{sf}^{Cap}$ and $\rho_{Cap}$ on the spin Hall torque. Our simulation suggests that a capping layer with a long $l_{sf}^{Cap}$ and high $\rho_{Cap}$ leads to an enhanced magnitude of spin Hall torque. Therefore, our model suggests the optimization of capping layer properties as an avenue to enhance the overall SOT, a key requirement in spin torque-based spintronic devices.

**ACKNOWLEDGMENTS**

This work is supported by the Singapore National Research Foundation (NRF), Prime Minister's Office, under its Competitive Research Programme (NRF CRP12-2013-01, NUS Grant No. R-263-000-B30-281), as well as the MOE-AcRF Tier-II grants: MOE2013-T2-2-125 (NUS Grant No. R-263-000-B10-112) and MOE2015-T2-1-099 (NUS Grant No. R-380-000-012-112).

**Appendix: Derivation of transverse spin current boundary conditions**

We introduce a "transmission matrix" $\hat{t}$ that gives the transmission probability of the source state $|s_a\rangle$ into the drain state $|d_b\rangle$, where $a$ and $b$ index the eigenspinor states at the source and drain segments, respectively. Explicitly, $\hat{t} = \sum_{a,b} |d_b\rangle t_{ba} \langle s_a|$, so that

$$\hat{t}\,|s_a\rangle = \sum_b |d_b\rangle t_{ba} \qquad (A1)$$

where $t_{ba}$ is the transmission probability from source state $a$ to the drain state $b$.

The density matrix at the drain due to state $|s_a\rangle$ being transmitted through the interface is given by $\rho_d^a = \hat{t}\,|s_a\rangle\langle s_a|\hat{t}^\dagger$. Let us consider the case of $|s_a\rangle = |+x\rangle$, a state perfectly polarized in the $+x$ direction. Then, $\rho_d^{+x} = \hat{t}\,|+x\rangle\langle +x|\hat{t}^\dagger$ can be thought of as the density matrix for the transmitted drain state corresponding to the source state polarized in the $+x$ direction.

More generally, the resulting drain density matrix $\rho_d$ due to the transmission of a mixture of source states corresponding to the source density matrix $\rho_s$ can be written as $\rho_d = \hat{t}\rho_s\hat{t}^\dagger$. For example, consider a source density matrix with populations $n_{\pm x}$ for the $+x$ and $-x$ polarized spins, respectively, i.e. $\rho_S = |+x\rangle n_+\langle +x| +$ $|+x\rangle n_-\langle +x|$. We then have $\rho_S = \frac{1}{2}((|+x\rangle\langle +x|+|-x\rangle\langle -x|)(n_+ + n_-) +$
$(|+x\rangle\langle +x|-|-x\rangle\langle -x|)(n_+ - n_-)) = \frac{1}{2}(I_2 n_q + \sigma_x n_x) \approx \frac{e}{2}(I_2 \mu_q + \sigma_x \mu_x)$. From this, we see that the electrochemical potential $\mu_a$ for the spin polarization $a$ is associated with the $\sigma_a$ operator. Both $\sigma_a$ and $\mu_a$ are related to the imbalance between the spins polarized along $+a$ and $-a$.

$\rho_d$ can be decomposed into components pointing in the various spin directions by using the prescription that for an arbitrary $(2 \times 2)$ matrix $\hat{m} = \sum_{i=\{I,x,y,z\}} \hat{\sigma}_i c_i$, we have $c_i = \frac{1}{2}\text{Tr}(\hat{m}\,\hat{\sigma}_i)$ where $\hat{\sigma}_I = I_2$. The resulting spin polarization $b$ in the drain segment due to a source spin of polarization $a$ incident on the interface is then $\rho_d^{ab} = \frac{1}{4}\text{Tr}(\hat{t}\hat{\sigma}_a \hat{t}^\dagger \hat{\sigma}_b)$.

The FM magnetization leads to anisotropic tunneling and reflection probabilities between charge carriers of different spin orientations, with the reference spin axis being along $\hat{m}$, where $\hat{m}$ denotes the FM magnetization direction. Thus, $\hat{t}$ takes the form of [25]

$$\hat{t} = t_0 I_2 + \delta t(\vec{\sigma} \cdot \hat{m}).$$

We set the spin $z$ direction to be along $\hat{m}$, so that $\hat{t} = |\uparrow\rangle t_\uparrow \langle\uparrow| + |\downarrow\rangle t_\downarrow \langle\downarrow|$ with $t_{\uparrow/\downarrow} = t_0 \pm \delta t$. (The $t_\uparrow$ here is in fact the $t_{\uparrow\uparrow}$ in Eq. A1 except that we have omitted the second up arrow for notational simplicity.) Considering the spin polarization $x$ in the drain segment, it then follows that

$$\begin{aligned}\text{Tr}(\hat{t}\,\hat{\sigma}_x \hat{t}^\dagger\,\hat{\sigma}_x) &= 2\text{Re}(t_\downarrow^* t_\uparrow);\\ \text{Tr}(\hat{t}\,\hat{\sigma}_y \hat{t}^\dagger\,\hat{\sigma}_x) &= -2\text{Im}(t_\uparrow^* t_\downarrow);\\ \text{Tr}(\hat{t}\,\hat{\sigma}_z \hat{t}^\dagger\,\hat{\sigma}_x) &= 0 \end{aligned} \quad (A2)$$

Let us define $G_{\uparrow\downarrow}^t \equiv \frac{e^2}{4\hbar} t_\uparrow^* t_\downarrow$, where the superscript $t$ stands for "transmission". Similarly, we can also define a corresponding $G_{\uparrow\downarrow}^r$ where we consider the states at the drain segment due to the *reflection* of drain states at the interface back to the drain segment. Thus, for a spin $x$ current at the drain segment, we have, by considering Eq. (A2), contributions from spin $x$ and $y$ states incident on both the source (transmitted contribution) and drain (reflected contribution) sides of the interface, giving $j_{s,x}|_d = 2\left(\text{Re}(G_{\uparrow\downarrow}^t)\mu_{s,x}|_s - \text{Im}(G_{\uparrow\downarrow}^t)\mu_{s,y}|_s + \text{Im}(G_{\uparrow\downarrow}^r)\mu_{s,y}|_d - \text{Re}(G_{\uparrow\downarrow}^r)\mu_{s,x}|_{s,d}\right)$.

Here, we define positive current to be flowing from the source to the drain direction. Thus, taking $G_{\uparrow\downarrow} > 0$, the reflection of drain states at the interface would yield a negative contribution to the current. Similarly, for the $y$ polarization in the drain, we have

$$\text{Tr}(\hat{t}\,\hat{\sigma}_x \hat{t}^\dagger\,\hat{\sigma}_y) = 2\text{Im}(t_\uparrow^* t_\downarrow);$$

$$\text{Tr}(\hat{t}\,\hat{\sigma}_y \hat{t}^\dagger\,\hat{\sigma}_y) = 2\text{Re}(t_\uparrow^* t_\downarrow);$$

$$\text{Tr}(\hat{t}\,\hat{\sigma}_z \hat{t}^\dagger\,\hat{\sigma}_y) = 0. \quad (A3)$$

Finally, the corresponding relations for the spin currents at the source side of the interface, i.e. $j_{s,x}|_s$ and $j_{s,y}|_s$ follow analogously.

## References

1. L. Liu, O. Lee, T. Gudmundsen, D. Ralph, and R. Buhrman, Physical review letters **109,** 096602 (2012).
2. I. M. Miron, K. Garello, G. Gaudin, P.-J. Zermatten, M. V. Costache, S. Auffret, S. Bandiera, B. Rodmacq, A. Schuhl, and P. Gambardella, Nature **476,** 189 (2011).
3. G. Siracusano, R. Tomasello, V. Puliafito, A. Giordano, B. Azzerboni, A. La Corte, M. Carpentieri, and G. Finocchio, Journal of Applied Physics **117,** 17E504 (2015).
4. S. Emori, U. Bauer, S.-M. Ahn, E. Martinez, and G. S. Beach, arXiv preprint arXiv:1302.2257 (2013).
5. P. Haazen and E. Mure, Nat. Mater **12,** 299 (2013).
6. P. M. Haney, H.-W. Lee, K.-J. Lee, A. Manchon, and M. D. Stiles, Physical Review B **87,** 174411 (2013).

7   A. Manchon, arXiv preprint arXiv:1204.4869 (2012).
8   J. Kim, J. Sinha, M. Hayashi, M. Yamanouchi, S. Fukami, T. Suzuki, S. Mitani, and H. Ohno, arXiv preprint arXiv:1207.2521 (2012).
9   R. Ramaswamy, X. Qiu, T. Dutta, S. D. Pollard, and H. Yang, Applied Physics Letters **108,** 202406 (2016).
10  X. Qiu, W. Legrand, P. He, Y. Wu, J. Yu, R. Ramaswamy, A. Manchon, and H. Yang, Physical review letters **117,** 217206 (2016).
11  J. Yu, X. Qiu, W. Legrand, and H. Yang, Applied Physics Letters **109,** 042403 (2016).
12  S. Woo, M. Mann, A. J. Tan, L. Caretta, and G. S. Beach, Applied Physics Letters **105,** 212404 (2014).
13  M. Morota, Y. Niimi, K. Ohnishi, D. Wei, T. Tanaka, H. Kontani, T. Kimura, and Y. Otani, Physical Review B **83,** 174405 (2011).
14  C.-F. Pai, L. Liu, Y. Li, H. Tseng, D. Ralph, and R. Buhrman, Applied Physics Letters **101,** 122404 (2012).
15  H. Wang, C. Du, Y. Pu, R. Adur, P. C. Hammel, and F. Yang, Physical review letters **112,** 197201 (2014).
16  T. Valet and A. Fert, Physical Review B **48,** 7099 (1993).
17  J. Barnaś, A. Fert, M. Gmitra, I. Weymann, and V. Dugaev, Physical Review B **72,** 024426 (2005).
18  N. Chung, M. Jalil, and S. Tan, Journal of Physics D: Applied Physics **42,** 195502 (2009).
19  M. Jalil, S. Tan, R. Law, and N. Chung, Journal of applied physics **101,** 124314 (2007).
20  S. G. Tan and M. B. Jalil, *Introduction to the Physics of Nanoelectronics* (Elsevier, 2012).
21  C. Petitjean, D. Luc, and X. Waintal, Physical review letters **109,** 117204 (2012).
22  V. Amin and M. Stiles, Physical Review B **94,** 104420 (2016).
23  S. Zhang, P. Levy, and A. Fert, Physical review letters **88,** 236601 (2002).
24  A. Shpiro, P. M. Levy, and S. Zhang, Physical Review B **67,** 104430 (2003).
25  A. Brataas, Y. V. Nazarov, and G. E. Bauer, The European Physical Journal B-Condensed Matter and Complex Systems **22,** 99 (2001).
26  S. Cho, S. H. Baek, K. D. Lee, Y. Jo, and B. G. Park, Sci Rep **5,** 14668 (2015).
27  Y. Jiang, S. Abe, T. Ochiai, T. Nozaki, A. Hirohata, N. Tezuka, and K. Inomata, Physical review letters **92,** 167204 (2004).
28  A. Ghosh, S. Auffret, U. Ebels, and W. Bailey, Physical review letters **109,** 127202 (2012).
29  L. Liu, C.-F. Pai, Y. Li, H. Tseng, D. Ralph, and R. Buhrman, Science **336,** 555 (2012).
30  P. He, X. Qiu, V. L. Zhang, Y. Wu, M. H. Kuok, and H. Yang, Advanced Electronic Materials **2** (2016).
31  J.-C. Rojas-Sánchez, N. Reyren, P. Laczkowski, W. Savero, J.-P. Attané, C. Deranlot, M. Jamet, J.-M. George, L. Vila, and H. Jaffrès, Physical review letters **112,** 106602 (2014).
32  W. Zhang, W. Han, X. Jiang, S.-H. Yang, and S. S. Parkin, Nature Physics **11,** 496 (2015).
33  V. P. Amin and M. D. Stiles, Physical Review B **94,** 104419 (2016).
34  K. Chen and S. Zhang, Physical review letters **114,** 126602 (2015).
35  K. D. Belashchenko, A. A. Kovalev, and M. van Schilfgaarde, Physical review letters **117,** 207204 (2016).